\documentclass[12pt]{article}
\usepackage{amsfonts,latexsym,fullpage,amsmath,latexsym,amssymb}

\newtheorem{Definition}{Definition}
\newtheorem{Theorem}{Theorem}
\newtheorem{Lemma}{Lemma}

\newcommand{\onemat}[0]{{\mathbf 1}}

\newcommand{\cI}[0]{{\mathcal I}}

\newcommand{\C}[0]{{\mathbb{C}}}

\newcommand{\qed}[0]{\hfill $\Box$}

\title{\Large \textbf{Lower Bound on the Chromatic Number by\\
Spectra of Weighted Adjacency Matrices}} \author{ Pawe{\l}
Wocjan\thanks{e-mail: {\protect\tt
\{wocjan,janzing,eiss\_office\}@ira.uka.de}}, Dominik Janzing, Thomas
Beth \\ \small Institut f{\"u}r Algorithmen und Kognitive Systeme,
Universit{\"a}t Karlsruhe,\\[-1ex] \small Am Fasanengarten 5,
D-76\,128 Karlsruhe, Germany\\[-1ex] \small Forschungsgruppe Quantum
Computing}

\date{December 21, 2001}

\begin{document}

\maketitle

\abstract{A lower bound on the chromatic number of a graph is derived
by majorization of spectra of weighted adjacency matrices. These
matrices are given by Hadamard products of the adjacency matrix and
arbitrary Hermitian matrices.}

\section{Introduction}
Let $G=(V,E)$ be a graph on $n$ vertices $V=\{0,\ldots,n-1\}$. An
admissible vertex coloring of $G$ is an assignment of colors to its
vertices such that no two adjacent vertices receive the same
color. The minimum number of colors needed for such a coloring is
called the {\it chromatic number} of $G$, and is denoted
$\chi$. Determining the chromatic number of a graph is known to be
NP-hard \cite{garey}. Besides its theoretical significance as a
canonical NP-hard problem, graph coloring arises naturally in a
variety of applications such as e.g. register allocation and
scheduling. Therefore it is important to have bounds on the chromatic
number.

Bounds are often computed with the help of the adjacency matrix. Let
$A=(a_{ij})$ denote the adjacency matrix of $G$, i.\,e.
$$
a_{kl}=\left\{
\begin{array}{ll}
1 & \mbox{if vertices $k$ and $l$ are adjacent} \\
0 & \mbox{otherwise}
\end{array}
\right.
$$
and $\lambda_1\ge\lambda_2\ge\cdots\ge\lambda_n$ denote the
eigenvalues of $A$. The upper bound $\lambda_1+1\ge\chi$ has been
shown in \cite{wilf}. The lower bound
\begin{equation}\label{hoffman}
\chi\ge\frac{\lambda_1}{|\lambda_n|}+1\,
\end{equation}
has been shown in \cite{Hoffman70}. The following Theorem proved in
\cite{barnes} gives a generalization of the bound (\ref{hoffman}).

\begin{Theorem}\label{barnestheorem}
Let $A$ denote the adjacency matrix for a connected graph $G$ on $n$
vertices, and let $D=diag(d_1,\ldots,d_n)$ be a diagonal matrix such
that $A+D$ is positive semidefinite. Then each $d_i$ is positive, and
the largest eigenvalue of the matrix $D^{-\frac{1}{2}}A
D^{-\frac{1}{2}}+I$ is a lower bound for $\chi$.
\end{Theorem}

Note that the bound (\ref{hoffman}) can be obtained from
Theorem~\ref{barnestheorem} by taking $D=|\lambda_n| I$.

\section{Lower bounds for $\chi$}
We derive a method to obtain lower bounds on the chromatic
number. This method contains the above results on the lower bound as
special cases.

First we define the problem to reverse the sign of a matrix and the
cost of doing this. We consider weighted adjacency matrices, i.e.\
special matrices constructed from the adjacency matrix of a graph $G$,
and show that $\chi-1$ is an upper bound on the cost. Then a lower
bound is derived on the cost in terms of the spectra of the weighted
adjacency matrices. This yields a lower bound on the chromatic number.
\begin{Definition}[Sign reversal]
Let $M\in\C^{n\times n}$ be any traceless matrix. A sign reversal map
$\cI:=(r_1,U_1;r_2,U_2;\ldots;r_N,U_N)$ for $M$ is characterized by
$N$ positive real numbers $r_j$ and unitary matrices
$U_j\in\C^{n\times n}$ such that
\begin{equation}
\sum_{j=1}^N r_j U_j^{-1} M U_j = -M\,.
\end{equation}
We call $c(\cI):=\sum_{j=1}^N r_j$ the cost of the sign reversal map
$\cI$. 
\end{Definition}
Using basic results of representation theory we can show that this is
always possible. Let $G$ be a group acting irreducibly on $\C^n$ via
the unitary representation $g\mapsto U_g$. Let $M\in\C^{n\times n}$
matrix. By Schur's Lemma \cite{group} we have
\begin{equation}
\frac{1}{|G|}\sum_{g\in G} U_g^{-1} M U_g = \frac{{\rm tr}(M)}{n}\onemat\,, 
\end{equation}
where ${\rm tr}(M)$ denotes the trace of $M$ and $\onemat$ is the
identity matrix of size $N$. 

In the following the matrix $M$ is always traceless. Then by conjugating
$M$ with all elements of $G$ but the identity we obtain 
\[
\sum_{G\setminus\{1\}} U_g M U_g = -M
\]
since $M$ is traceless.

We consider matrices constructed from the adjacency matrix of a graph
$G$ and derive a lower bound on the cost of sign reversal in terms of
the chromatic number of $G$. To do this we need the definition of the
Hadamard product. If $A=(a_{ij})\in\C^{n\times n}$ and
$B=(b_{ij})\in\C^{n\times n}$ are given, then the Hadamard product of
$A$ and $B$ is the matrix $A*B=(a_{ij} b_{ij})\in\C^{n\times n}$
\cite{HJ90}.

\begin{Lemma}\label{invertChromatic}
Let $A=(a_{kl})$ be the adjacency matrix of a graph $G=(V,E)$ with $n$
vertices and chromatic number $\chi$. For an arbitrary matrix
$W=(w_{kl})\in\C^{n\times n}$ the sign of the Hadamard product
$M:=W*A$ can be reversed with cost $\chi-1$.
\end{Lemma}
{\bf Proof:} Choose a partition (disjoint union) of the vertices
$V=V_0\cup V_1\cup\ldots\cup V_{\chi-1}$ corresponding to a (minimal)
coloring. The set $V_c$ contains all vertices of color $c$
($c\in\{0,\ldots,\chi-1\})$. We assume that the vertices are ordered
according to their colors, i.e. first come the vertices of color $0$,
than of color $1$, etc.

Let $\omega\in\mathbb{C}$ be a primitive $\chi$-th root of unity,
i.e. $\omega^{\chi}=1$. Define a diagonal matrix $D:={\rm
diag}(d_0,d_1,\ldots,d_{n-1})$ where $d_i:=\omega^c$ if $i\in V_c$. We
show that
\[
\bar{M}:=\sum_{j=0}^{\chi-1} D^{-j} M D^j\,.
\]
is the zero matrix. Let $M=(m_{kl})_{k,l=0,\ldots n-1}$ and
$\bar{M}=(\bar{m}_{kl})_{k,l=0,\ldots n-1}$. The entries of $\bar{M}$
are given by
\[
\bar{m}_{kl}=\sum_{j=0}^{\chi-1} \omega^{-kj} m_{kl} \omega^{lj}\,.
\]
Let $k,l\in V_c$ with $k\neq l$. The vertices $k$ and $l$ have the the
same color and consequently they cannot be adjacent,
i.e. $a_{kl}=0$. Therefore we have $\bar{m}_{kl}$ since $m_{kl}=w_{kl}
a_{kl}=0$. Note that $\bar{m}_{kk}=0$ since $a_{kk}=0$ (the diagonal
entries of the adjacency matrix are all zero).

Now let $k\in V_c$ and $l\in V_{\tilde{c}}$ with $c\neq\tilde{c}$. We
have
\[
\bar{m}_{kl}=\sum_{j=0}^{\chi-1} \omega^{-kj} m_{kl} \omega^{lj}=0
\]
since the vectors $({\omega}^{c\, 0},{\omega}^{c\, 1},\ldots,
{\omega}^{c\,(\chi-1)})$ and $({\omega}^{\tilde{c}\,
0},{\omega}^{\tilde{c}\, 1},\ldots, {\omega}^{\tilde{c}\, (\chi-1)})$
are orthogonal. Note that they are rows of the matrix of the discrete
Fourier transform of size $\chi$.  \qed

\vspace*{12pt} To derive a lower bound on the cost of a sign reversal
map we introduce the concept that is known as spectral majorization of
matrices \cite{Bhatia:96}. Let $x=(x_1,\ldots,x_n)$ and
$y=(y_1,\ldots,y_n)$ be two $n$-dimensional real vectors. We introduce
the notation $\downarrow$ to denote the components of a vector
rearranged into non-increasing order, so
$x^\downarrow=(x_1^\downarrow,\ldots,x_n^\downarrow)$, where
$x_1^\downarrow\ge x_2^\downarrow\ge\ldots\ge x_n^\downarrow$. We say
that $x$ is majorized by $y$ and write $x\prec y$, if
\[
\sum_{i=1}^m x_i^\downarrow \le \sum_{i=1}^m y_i^\downarrow\,,
\]
for $m=1,\ldots,n-1$ and $\sum_{i=1}^n x_i^\downarrow = \sum_{i=1}^n
y_i^\downarrow$.

Let $\mathrm{Spec}(X)$ denote the spectrum of the Hermitian matrix
$X$, i.e.\ the vector of eigenvalues. Recall that the eigenvalues of a
Hermitian matrix are real. Ky Fan's maximum principle \cite{Bhatia:96}
gives rise to a useful constraint on the eigenvalues of a sum of
Hermitian matrices $A_j$ for $j=1,\ldots,N$:
\begin{equation}\label{major}
\mathrm{Spec}\big(\sum_{j=1}^N A_j\big)\prec\sum_{j=1}^N
\mathrm{Spec}(A_j)\,.
\end{equation}
The following lemma gives a lower bound on the cost of sign reversal
in terms of the spectrum of $M$.
\begin{Lemma}\label{lowerBound}
A lower bound on the negation cost of a Hermitian matrix $M$ is given
by the smallest positive real number $\tau$ such that
\begin{equation}
\mathrm{Spec}(-M)\prec\tau\,\mathrm{Spec}(M)\,.
\end{equation}
\end{Lemma}
{\bf Proof:} Let $\cI:=(r_1,U_1;r_2,U_2;\ldots r_N,U_N)$ be an
arbitrary sign reversal map for $M$, i.e.
\begin{equation}\label{inverteq}
\sum_{j=1}^N r_j\, U_j^{-1} M U_j = -M\,.
\end{equation}
Applying the inequality~(\ref{major}) to eq.~(\ref{inverteq}) yields
\[
\mathrm{Spec}(-M)=\mathrm{Spec}\big(\sum_{j=1}^N r_j U_j^{-1} M
U_j\big) \prec \sum_{j=1}^N r_j\,\mathrm{Spec}(U_j^{-1} M U_j) =
c(\cI)\,\mathrm{Spec}(M)\,.
\]
Therefore for all sign reversal maps we have $c(\cI)\ge\tau$.  \qed

\vspace*{12pt} Let $\lambda_1,\lambda_2,\ldots,\lambda_n$ be the
eigenvalues of $M$ with
$\lambda_1\ge\lambda_2\ge\ldots\ge\lambda_n$. Then the lower bound $\tau$
is given by
\begin{equation}\label{max}
\tau:=\max_{m=1,\ldots n-1}\Big\{ \frac{\sum_{i=1}^m \lambda_i}{-\sum_{i=1}^m
\lambda_{n+1-i}}\Big\}
\end{equation}
since for the eigenvalues $\mu_i$ of $-M$ we have
$\mu_i=\lambda_{n+1-i}$ ($i=1\ldots,n$).

By using the results on the lower and upper bounds on the cost of sign
reversal we obtain:
\begin{Theorem}[Lower bound on the chromatic number]
Let $A$ be the adjacency matrix of a graph $G$ with $n$ vertices and
$\chi$ its chromatic number. For any Hermitian matrix $W\in\C^{n\times
n}$ let $\tau_W$ denote the minimal positive real number such that
\begin{equation}\label{lowerW}
\lambda(-W*A)\prec\tau_W\,\lambda(W*A)\,.
\end{equation}
Let $\tau$ be the maximal $\tau_W$, where the maximum is taken over
all Hermitian matrices $W\in\C^{n\times n}$. Then we have
\begin{equation}
\chi \ge \tau+1\,.
\end{equation}
\end{Theorem}
{\bf Proof:} Lemma~\ref{invertChromatic} shows that the sign of all
matrices of the form $W*A$ can be reversed with cost
$\chi-1$. Lemma~\ref{lowerBound} gives $\tau_W$ in eq.~(\ref{lowerW})
as a lower bound on the sign reversal cost of $W*A$. Consequently, we
have $\chi\ge \tau_W +1$ for all Hermitian matrices. We choose the
maximal $\tau_W$ to obtain the best bound on $\chi$. \qed

\vspace*{12pt} Note that we obtain as a special case the well-known
lower bound $\chi\ge\frac{\lambda_1}{|\lambda_n|}+1$. Set $W$ to be
the matrix all of whose entries are equal to $1$. Then we have
$A=W*A$. Instead of taking the maximum over $m=1,\ldots,n-1$ consider
only $m:=1$ in eq.~(\ref{max}). This corresponds to the maximal and
minimal eigenvalues of $A$.

We now show that Theorem~1 can also be understood as a special case of
Theorem~2. Let $D:=diag(d_0,\ldots,d_{n-1})$ be the diagonal matrix
defined as in Theorem~1. Take $W$ to be the matrix with entries
$w_{kl}:=\sqrt{d_k}\sqrt{d_l}$. Then we have $W*A=D^{-\frac{1}{2}} A
D^{-\frac{1}{2}}$. This shows that this modified adjacency matrix can
also be expressed with the Hadamard product.

As a conclusion Theorem~2 permits to consider a larger class of
modified adjacency matrices and to take into account all eigenvalues
(Theorem~1 considers only the maximal and minimal eigenvalues). The
advantage of the method presented in \cite{barnes} is that the matrix
$D$ is the solution of a semidefinite programming problem that can be
computed by an algorithm given there. It remains to be shown whether
the larger class of modified adjacency matrices permits to obtain
better lower bounds and whether there are efficient algorithms to
compute them. Nevertheless, the derivation in our approach is easy and
gives a generalization of Theorem~2.

We would like to point out that the sign reversal map described here
is an abstraction of a problem in quantum physics. A physical system
evolving due to the Schr\"odinger equation
\[
\frac{d}{dt}
|\psi (t)\rangle = -i H |\psi (t)\rangle
\]
(where $H$ is the so-called ``Hamilton operator'') can be made evolve
backwards in time by interspersing the natural time evolution with
external control operations \cite{graph,arrow}. The unitary maps $U_j$
in Definition~1 are abstractions of control operations. The vertices
of the graph represent particles and the edges indicate whether there
is an interaction between the two particles or not. The physical
meaning of the inversion cost is the time overhead for simulating the
inverse evolution, i.e., the factor by which the reverse dynamics is
slower than the original one. The cost is therefore the {\it
complexity} of reversing the dynamical evolution and the chromatic
number is an upper bound on this time complexity.

\section*{Acknowledgments}
This work has been supported by grands of the DFG Schwerpunktprogramm
{\it Komplexit\"at und Energie}.

%%%%%%%%%%%%%%%%%%%%%%%%%%%%%%%%%%%%%%%%%%%%%%%%%%%%%%%%%%%%
%
% The Literature
%
%%%%%%%%%%%%%%%%%%%%%%%%%%%%%%%%%%%%%%%%%%%%%%%%%%%%%%%%%%%%

\end{document}